# Towards a dynamically reconfigurable pixelated reflective display: Focused ion beam for phase-change metapixel structures

Daniel T. Yimam*, Minpeng Liang, Jianting Ye, Bart J. Kooi*

Zernike Institute for Advanced Materials, University of Groningen, Nijenborgh 4, 9747 AG Groningen, The Netherlands

**Abstract**

The switching and optical properties of phase-change thin films are actively investigated for future smart optical devices. The possibility of having more than one stable state, the large optical contrast between phases, and the fast and reversible switching are some attractive properties driving the research interest. Optical devices based on phase change alloys are considered the frontier contenders for tunable photonics. The combination of vivid structural color formation, with partial amorphization/crystallization of phase change alloys, and the associated optical tunability could be integrated into an energy-efficient reflective display device with high pixel density. This work demonstrates a contrast formation due to relative height differences from isolated pixelated structures. A reflective heterostructure device consisting of a low-loss $Sb_2Se_3$ alloy on a gold substrate was produced. With a focused ion beam, a pixelated metasurface structure was produced. Moreover, the ability to create local height differences using an ion beam was employed to create a structural color combination mimicking traditional LED like RGB pixels. We believe our approach in creating metapixels on phase change thin film surfaces could open up research interest in phase change alloys and moving away from semi/static plasmonic systems into truly dynamic display devices.

# Introduction

The demand for display devices has multiplied manifold over the past few decades with the application of visual information transfer coupled with fast data communications.[1] From cathode ray tubes (CRT) to liquid crystal displays (LCD) and light emitting diodes (LED), pixelated display devices push forward the flow of information communication with attractive colors and appealing visual contents.[1–3] Over the past few years, research has moved away from the traditional emission-based display devices towards reflective-based display devices with little or no internal light source.[1,4,5] Two main apparent reasons for the research interest and shift towards reflective-based display devices is the relatively low power consumption[4,6] and the possibility of moving towards large pixel-per-inch (PPI) values with high-resolution image formations.[7–9]

Reflective structural colors are a primary contender for high-resolution printing and dynamic display devices.[2,3] The most intense research in achieving structural colors for such applications involved plasmonic nanostructures and artificial metasurfaces.[9–14] Such artificially produced nanopillars and reflective structures provide an opportunity to produce tunable and intense reflectance intensity and high-resolution images in relatively small dimensions.[2,3] Research works on pixelated surfaces, based on plasmonic nanostructures, have reported the ease of reflectance control by shape and size changes of the resonators and nanopillars. Although amazing progress has been shown in structural colors-based nanoprinting and contrast formation, most results are static. The 'dynamic' aspect is limited to light polarization[15,16], chemical changes[17], and materials phase-switching.[18–20] A move from static/semi-static images towards a dynamically reconfigurable device structure with cyclability and active color control modes still has a long way to go. Phase-change materials (PCMs), with more than one stable phase with significant optical contrast and reversible switching, received broad interest for dynamically reconfigurable photonics applications.[21,22] The reflectance tuning, thus structural colors changes, have previously been shown by utilizing the phase-switching of PCMs by an external stimulus like laser and resistive heating.[23,24] However, the full power of PCMs lies in their ability to partially amorphize/crystallize locally. This partial structural transformation of PCMs corresponds to forming multiple reflectance states, thus structural colors.[19,20,25,26]

Here, we used a focused ion beam (FIB) to nanostructure isolated pixelated structures on a reflective heterostructure device with an active PCM material. A reflective device based on the strong-interference formalism was produced using pulsed laser deposition (PLD), where a low-loss $Sb_2Se_3$ PCM thin film was deposited on a reflective gold substrate. We previously showed how FIB's milling/sputtering power could be precisely controlled to produce sub-nm local depth variations which translated in structural color variations[27]. Our work showed the possibility of creating local structural colors of any dimension by a controlled ion milling/sputtering of a reflective heterostructure. The structural color variation was due to the relative depth change produced by the focused ion beam and could be the basis for creating metasurfaces for display applications. In the present work, we explicitly introduce pixels, i.e. RGB values of digital images are translated into isolated pixelated structures with local depth variations creating contrast. Our innovative approaches of integrating FIB for milling pixelated structures on a reflective device provide an opportunity to control both lateral and height dimensions, which currently is not possible with traditional lithography techniques. Moreover, we produced a metapixel surface with multiple structural color combinations imitating the conventional RGB display devices like LED displays. The optical tunability of phase change alloys can be used to locally tune the reflectance of individual pixels accounting for the final weighted value. In contrast, using nanoresonators and nanoantennas for creating colors is more complicated, also due to the plasmonic coupling between adjusted structures contributing to the weighted outcome.[28] FIB offers flexibility and additional degrees of freedom for effective integration of metasurfaces. A PCM-based device with local reflectance tunability from partial (amorphous-crystalline) phase transformations, could be a move forward to a truly dynamic display device.

**Experimental**

A reflective heterostructure device was prepared by pulsed laser deposition (PLD) of a phase change alloy $Sb_2Se_3$ on a gold substrate. A 1 $Jcm^{-1}$ laser fluence, 1 Hz repetition rate, and 0.12 mbar of Ar processing gas were used for the deposition. A powder-sintered target of $Sb_2Se_3$ was used for ablation. All depositions were done at room temperature, producing an as-deposited amorphous $Sb_2Se_3$ thin film. The gold substrate was deposited on a $Si/SiO_2$ substrate using the Temescal FC2000 electron beam evaporator. Before the deposition of $Sb_2Se_3$ on the gold substrate, a 10 nm spacer $LaAlO_x$ (LAO) layer was deposited by PLD to avoid intermixing. Before producing the reflective heterostructure by PLD, the reflectance profiles were calculated using the transfer-matrix algorithm. The optical constants n and k of individual layers in the heterostructure were first extracted using variable-angle spectroscopic ellipsometry. Then the extracted optical constants were used for reflectance calculations. The incidence angle-dependent and polarization-dependent reflectance profiles were collected using the same spectroscopic ellipsometry setup in reflection/transmission mode. The normal incidence reflectance profiles of locally produced structural colors (using FIB) were collected using a home-built system to collect spectra from <10 μm regions.

The metasufaces of individual isolated pixels with depth variations and the metapixelated surface were prepared by a focused ion beam (FIB of FEI Helios G4 CX Dual Beam system) operating at 30 kV accelerating voltage. MatLab was first used to process digital images before being loaded into the FIB software. Then, the milling/sputtering power of the FIB was calibrated to the grayscale values of the loaded digital images. The morphology of the produced surface structures was first imaged by scanning electron microscopy (SEM, Helios G4 CX) and an optical microscope. More information about optical, structural, and morphological analysis of similar reflective heterostructure devices and the FIB milling/sputtering calibration can be found in our previous work[27], to which we refer for additional details.

## Results and discussion

The strong interference of a thin film coating directly on a reflective substrate could, in principle, provide an advantage over other approaches like the Fabry-Perot cavity. The robust reflectance tuning with film thickness changes and the reduced path length from the reduced layer thicknesses promotes little phase accumulation and produces robust structures to incidence angle changes.[29] We produced a reflective heterostructure device based on the strong interference formalism, where a PCM layer of $Sb_2Se_3$ (typically 25 nm thick) is pulsed laser deposited on a reflective gold substrate. An LaAlOx (LAO) layer of 10 nm thick was used as a spacer layer to avoid intermixing between the active $Sb_2Se_3$ layer and the gold substrate. The reflectance spectra of the s- and p- polarized lights for the as-deposited amorphous phase of the 25 nm thick $Sb_2Se_3$ film are presented in Fig. 1. The reflectance calculations were done by employing the transfer-matrix formalism with a home written script. We used spectroscopic ellipsometry (J. Woollam UV-VIS) capable of angle-dependent reflectance measurements for the polarisation-dependent reflectance measurements. From Fig. 1, the measured s- and p-polarization reflectance spectra match very well with the calculated values. For both the s- and p- polarizations, the reflectance spectra show consistent profiles with little change for most of the incidence angles (up to ≈60º). This is expected for our reflective heterostructure device with a relatively small thickness compared to the wavelength of the light.

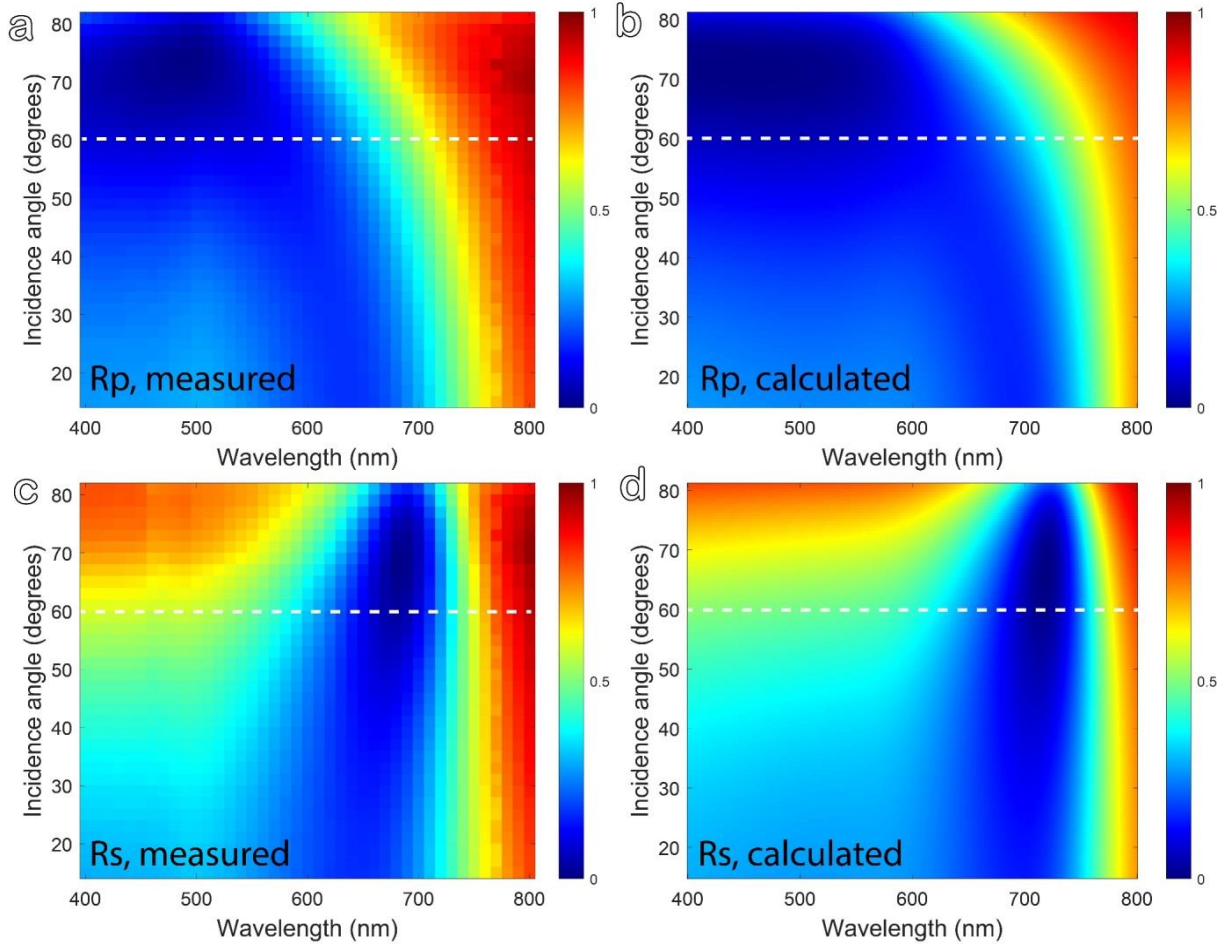

**Figure 1.** Experimental and calculated s- and p- polarization reflectance spectra for a reflective heterostructure device comprise of a 25 nm thick phase-change material $Sb_2Se_3$ (as-deposited amorphous phase) on a reflective gold substrate. Experimental and simulated reflectance for (a) and (b) p-polarized and (c) and (d) s-polarized light. The reflectance was measured and calculated for incidence angles of 15 to 80 degrees.

Structural colors provide an excellent opportunity to move toward an ultrahigh-resolution display device. Creating small structural details or 'pixels' with tunable reflectance states could be a way forward for a dynamic display device. As discussed above, the ability to fully tune reflectance with thickness and the limited reflectance deviation with the angle of incidence make the structures attractive. However, to fully utilize the pixelated structures of the reflective device, we first have to realize the contrast formation based on local pixelated structures. We previously reported a contrast formation and ultrahigh-resolution nanoprinting on reflective surfaces by employing a focused ion beam (FIB)[27]. Here, we extend the capabilities of milling/sputtering material away from the reflective device by FIB to locally produce pixelated structures with much higher resolutions than reported in our previous work. To move towards a dynamically reconfigurable device, we should create individual and isolated pixels with the possibility of changing their reflectance states individually.

The first logical innovation with respect to our previous work is to explicitly introduce pixels, but each pixel is not the combined effect of three pixels (like Red-Green-Blue (RGB) or Cyan-Magenta-Yellow (CMY)). Here, the milling/sputtering was done such that the final produced structure consists of isolated 'pillar-like' structures with the same lateral dimensions but different depths. Remember that although we used the word 'pillar' to describe the individually isolated pixels, they are, in principle, perforated regions with different depth values. Moreover, it is worth mentioning that the ability to create pixels with a continuous range of height variations is one of the advantages of using FIB and this is not possible with other techniques, e.g. based on lithography. Here, we demonstrate our approach using a 24-bit image of the painting *Starry Night* by Vincent van Gogh as an example. In a first step this image is converted into an 8-bit gray scale image that is then loaded into the FIB software for milling/sputtering material away based on the grayscale value–milling power relation. More explanation of the grayscale value and removed material calibration can be found in our previous work[27]. An SEM image of the resulting structured surface as produced by backscattered electrons (BSEs) is shown in Fig. 2a. Since BSEs scale with atomic number Z, thin $Sb_2Se_3$ regions closer to the gold substrate appear brighter than thick $Sb_2Se_3$ regions which stayed closer to the surface. The local height difference is visible from the contrast in the BSEs-SEM image resembling the original digital image. The surface is made up of 400 X 320 isolated pixel-like structures in a 100 X 80 $\mu m^2$ area. Notice the incredibly detailed features produced from the relatively small isolated pixels and their local changes in depth. To have a localized look at the isolated pixels, we zoom into some regions of the structure, as shown in Fig. 2b – 2e. Two regions, labelled *i* and *ii* and indicated by cyan and blue boxes, are selected for better visualization and presented in Figures 2b and 2c. A periodic array of pixels with different contrasts are visible in both images. For a region indicated by a red box in Fig. 2b, an even closer look at the periodic pixelated arrays is given in Fig. 2d and 2e. Fig. 2e is constructed with an artificial color map (Viridis color map) to visualize the height difference between individual pixelated structures from the visible intensity contrast between them.

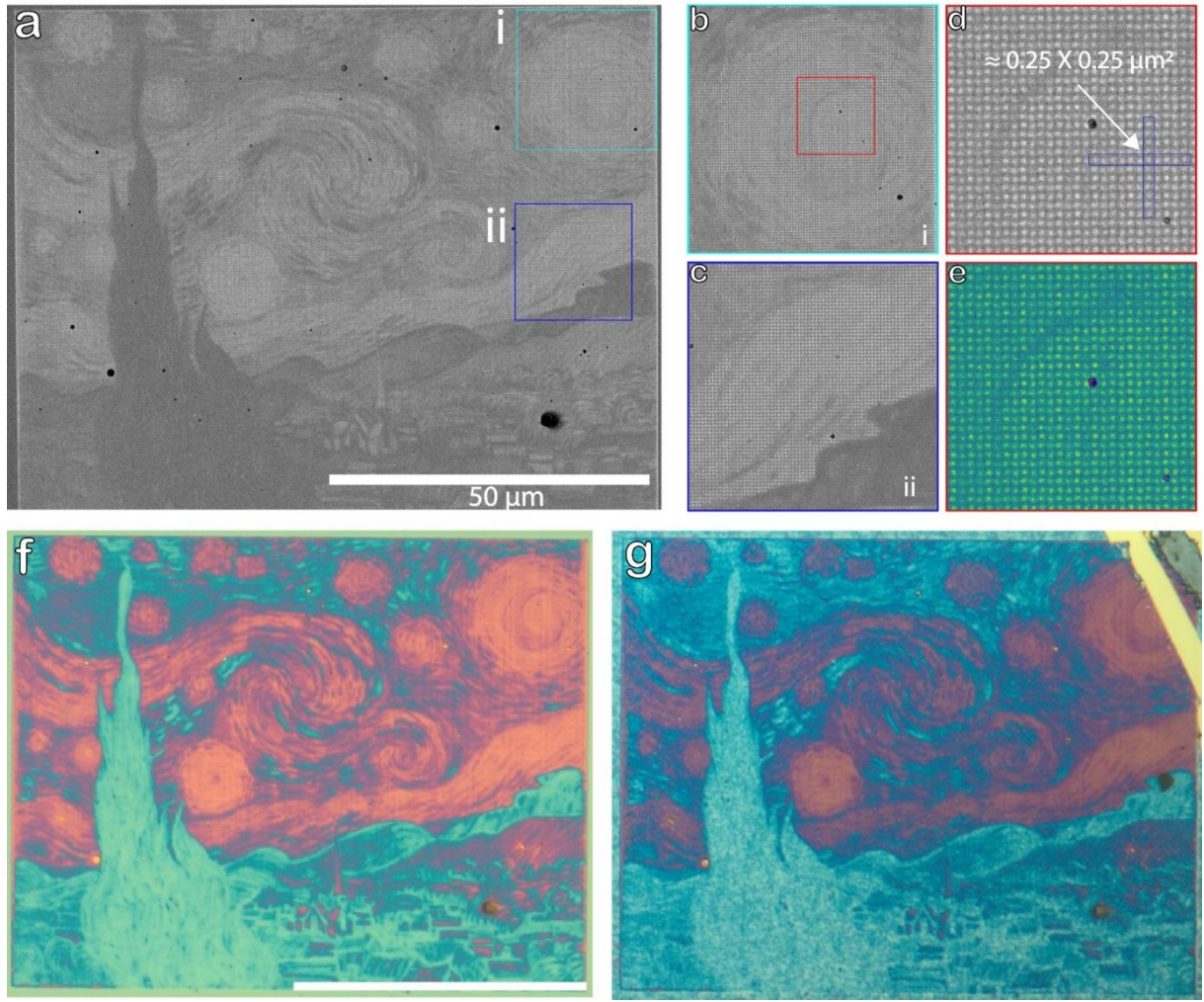

**Figure 2.** Pixelated structures produced by FIB. The painting *Starry Night* by Vincent van Gogh is used to produce the pixelated structures. (a) An SEM image of the structured surface produced by FIB. The BSEs signal formed the SEM image with local contrast due to the depth variation of individual 'pixel' squares. Closer looks at some regions of the pixelated surface are presented in (b) – (e). Individual and isolated pixel-like squares with varying depths are given in (d), and the Viridis color map is in (e) for better visualization. The optical images of the produced structure for the as-deposited amorphous and crystalline phases of the $Sb_2Se_3$ layer are presented in (f) and (g), respectively.

The created arrays of pixel-like structures are presented in Fig. 2d and 2e, where the separation between the pixels can be directly observed. Individually structured regions, i.e. pixels, cover an area of only 0.25 X 0.25 μm$^2$. Note that the dimensions choice is within a large range arbitrary, and FIB provides flexibility in tuning for the desired application. The intensity difference between individual structures can be seen from the presented results in Fig. 2d and the Viridis color map in Fig. 2e. This intensity difference is the cause of the contrast created in Fig. 2a and is caused by the relative depth difference between them. The height differences translate further into color contrasts. The optical images of the surface structure for both the as-deposited amorphous and crystalline phases of the $Sb_2Se_3$ film are presented in Fig. 2f and 2g, respectively. Indeed, the height difference between the isolated individual pixel-like structures produce vivid colors with local contrasts similar to the original digital image. Furthermore, the crystallization of the active PCM $Sb_2Se_3$ film produced another set of colors creating additional contrast. In supplementary information, additional examples of optical contrast formation from isolated individual pixel-like structures are given (see SI Fig. Sxx).

In a next step we want to retain the RGB values of a 24-bit image. Again, the painting *Starry Night* by Vincent van Gogh was taken as example. In the original image each pixel has a specific set of RGB values, but each pixel in Fig. 3a is now split up in three column pixels (as seen in the lower right of Fig. 3a) such that these three pixels together form a square pixel like in the original image. The left column pixel present R(ed), the middle column pixel G(reen) and the right column pixel B(lue). The idea here is to 'imitate' a traditional RGB pixelated display. The zoomed-in images in Fig. 3b and 3c clearly show the presence of the columns of pixels in the converted digital image. Now the individual RGB column pixels values are translated into 8-bit gray values. So, in Fig. 2 the 24-bit image was transformed directly into an 8-bit image as input for FIB, but now in Fig. 3 the 24-bits are kind of preserved. The produced digital image is then loaded into the FIB software for milling/sputtering material away based on the grayscale value–milling power relation. SEM images of the structured surface, based on backscattered (BSEs) and secondary (SEs) electron signals, are presented in Fig. 3d and 3e, respectively, and the zoomed-in images in Fig. 3f and 3g.

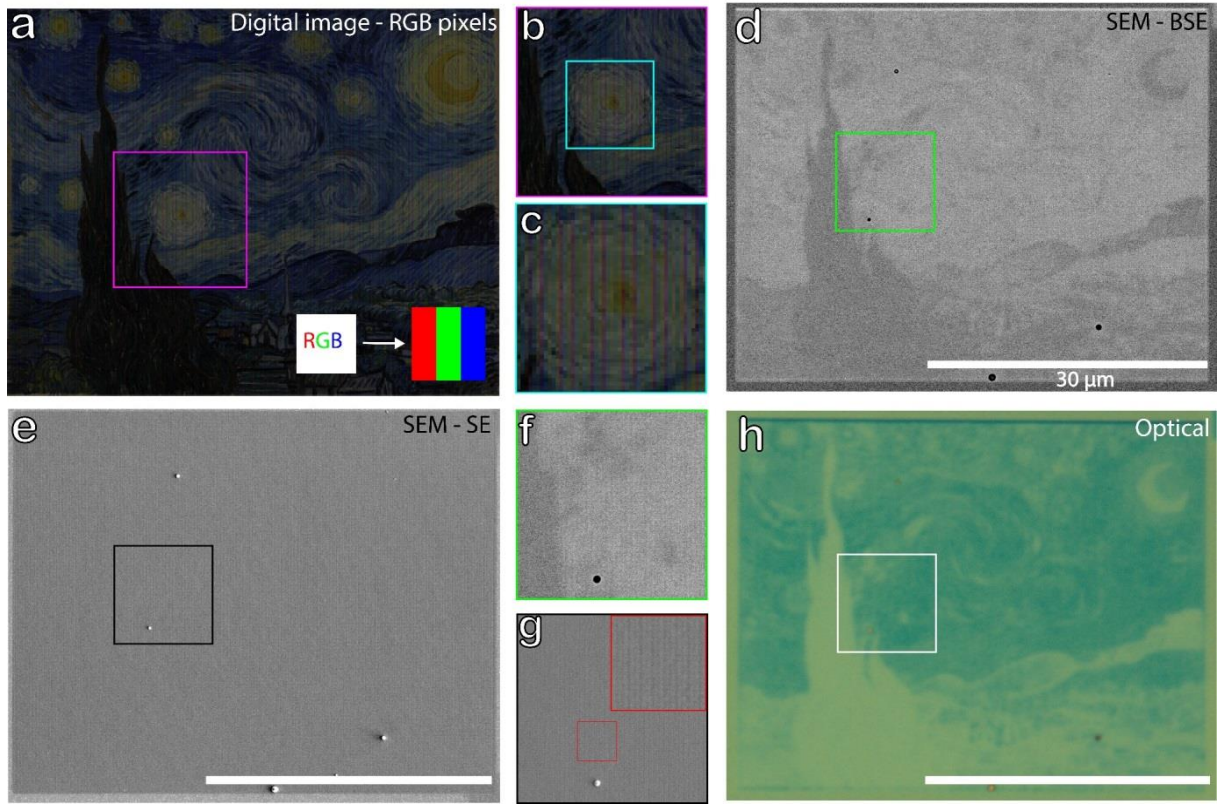

**Figure 3.** Focused ion beam milling of a reflective device for contrast formation imitating pixelated display view. (a) A digital image of the painting *Starry Night* by Vincent van Gogh is used to produce the structured surface. Each square pixel of the original 24-bit image was split up into 3 rectangular (columnar) pixels to represent the separate RGB values. A zoomed-in view of the pixelated digital image is given in (b) and (c). A column of RGB pixels is seen, reproducing the original single pixel. The original digital image was loaded into the FIB software and was used to mill/sputter away material from the surface of the reflective device. An SEM image created from (d) backscattered electrons (BSEs) and (e) secondary electrons (SEs) signal of the structure produced by using the digital image in (a). In (f) and (g) a closer look at the structure produced and the contrast formed are presented for both the BSEs and SEs signal SEM images. The optical image of the milled structure is presented in (h). High contrast with details similar to the original digital image is formed.

The two types of SEM images show interesting features about the structured surface. 1) The SEM image created by the BSEs shows contrast that scales with the local atomic number Z in the region imaged. So, thin $Sb_2Se_3$ regions closer to the gold substrate appear brighter than thick $Sb_2Se_3$ regions which stayed closer to the surface. The BSEs signal change with $Sb_2Se_3$ film thickness change is visible in Fig. 3d and creates a contrast similar to the uploaded digital image. 2) The SEM image created by SEs are more surface sensitive and show the surface morphology of the structured surface. The overview image Fig. 3e is rather boring and this is good news. Its shows that the strong contrast in BSE image is not reproduced in the SE image and that surface roughness is not a relevant factor. Still, when zooming in the topographic information in SE mode highlights the underlying pixelated structure. Indeed, taking a closer view in Fig. 3e and 3g shows the columnar structures formed similar to the digital image used for milling. From the complementary SEM images, based on BSEs signal for depth variation and SEs signals for topography, the formation of local columnar 'pixelated' structures from the FIB-device interaction can be seen clearly. The columnar structures and the height variations between them should translate to different contrast. This contrast is indeed visible in the optical image of the structure as shown in Fig. 3h, where an ultrahigh-resolution image similar to the original digital image is formed. This is indisputable proof that structural colors-based pixelated structures can produce vivid contrasts.

Although we showed optical contrast formation from 'pixelated' structures in Fig. 2 and 3, the dynamic reconfigurability is limited to nanoprinting a specific digital image on the surface and not yet exploits another degree of freedom for contrast formation from the phase switching of the active PCM thin film. This also holds for previously reported results in this research field, where individual plasmonic nano-pillars or nano-antennas are considered for pixelated structures creating static images with limited dynamic color tuning from phase switching or light polarization-dependent reflectance. To truly move towards a dynamically reconfigurable display device, we have to produce a structure capable of accommodating the required color contrast and dynamic range. For example, color formation by adding primary colors can be a suitable alternative, given how easily we can produce structural colors. Here, we propose a design consideration for a dynamically reconfigurable display device based on combined pixelated structures. Fig. 4a shows a schematic description of a possible structural color combination producing a 'pixelated' system similar to the widely used current RGB display devices based for instance on LED. Note, however, that here we produce a display fully based on reflection of light and therefore it does not require any power for light generation. Since the relatively small thickness change in the strong interference formalism is associated with a significant reflectance change and thus different structural colors, we can place multiple blocks of different depths next to each other. To realize all accessible structural colors throughout the thickness of the $Sb_2Se_3$ film, we systematically milled/sputtered local regions of the layer by varying the milling power of the FIB. The optical image in Fig. 4b shows various locally created structural colors with varying depths. This ability of FIB to create reflectance states with great lateral and vertical dimension controls provides an opportunity to create a combination of 'pixel blocks' with different structural colors. As a proof of concept, we selected three colors, labelled *i*, *ii*, and *iii* in Fig. 4b, to be combined in a pixelated device. The three selected colors for the as-deposited amorphous and crystalline phases of the $Sb_2Se_3$ film are shown in Fig. 4c. In addition, the measured reflectance spectra for the selected structural colors are shown in Fig. 4d, with a clear cavity mode translation. Note the difference in the structural colors and the reflectance states between the as-deposited amorphous and crystalline phases of the $Sb_2Se_3$ film, which is crucial for the proposed design. More on this below.

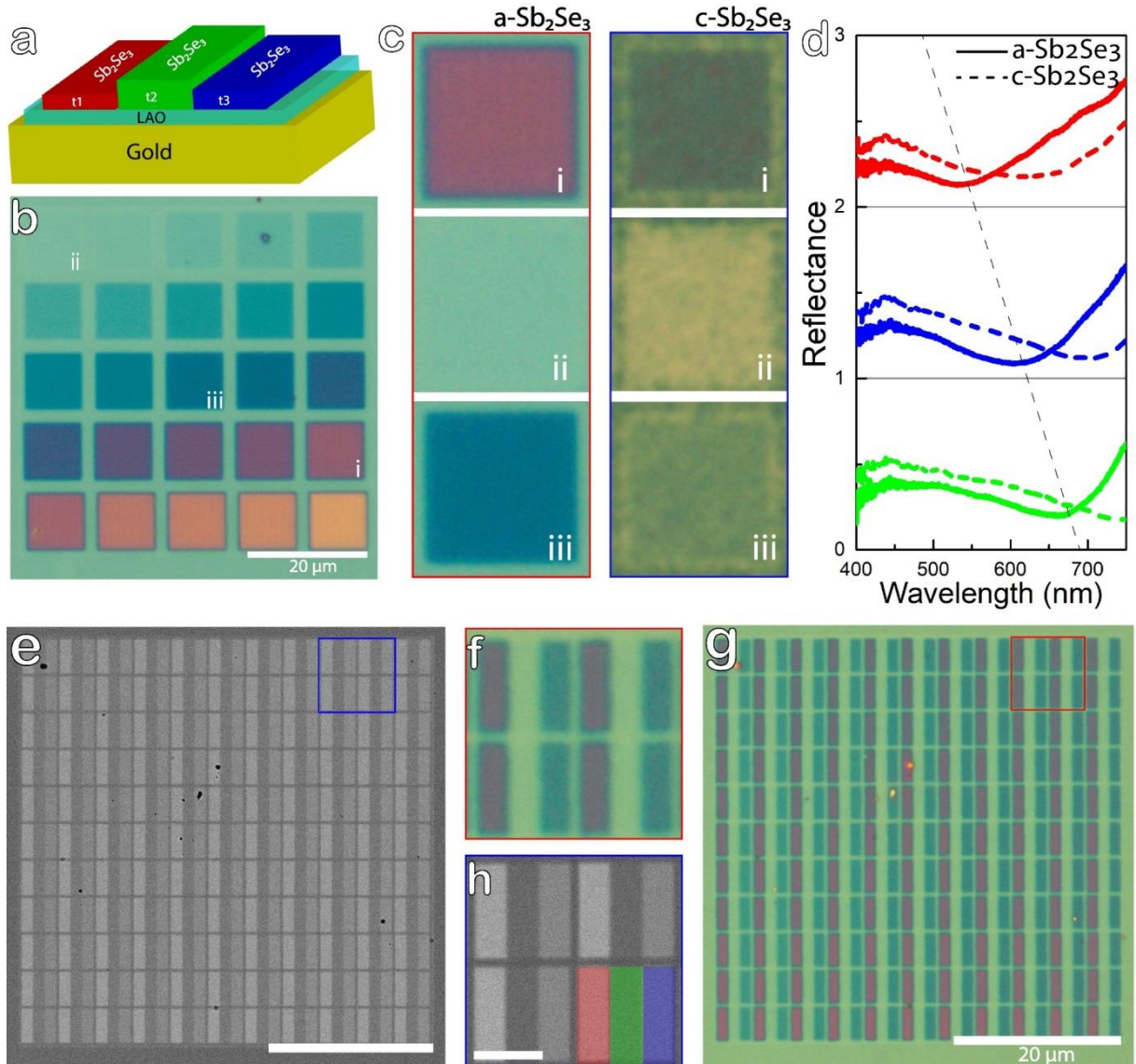

**Figure 4.** Structural colors based pixel formation. (a) A schematic of a reflective heterostructure device with structures of different heights stacked together. The heterostructure device consists of a reflective gold substrate, a phase-change material $Sb_2Se_3$ on top, and a thin spacer LAO layer. (b) Multiple structural colors are produced by locally changing the thickness of the active $Sb_2Se_3$ using FIB. A grayscale value of a digital image uploaded into the FIB software was used to control the depth of each square. Three structural colors (i – iii), from all the accessible colors seen in (b), were chosen to produce structural color-based pixel combinations as a proof of concept. The three colors selected are presented in (c) for the as-deposited amorphous and crystalline phases of the $Sb_2Se_3$ layer. (c) The measured reflectivity spectra of the three selected structural colors in (b) for the as-deposited amorphous and crystalline $Sb_2Se_3$ layer. Multiple 'pixels' on a row were produced by combining the three chosen structural colors in (b). An SEM image of the produced pixel-like structure and a closer look into the part of the structure are presented in (e) and (h), respectively. The optical image of the produced pixel-like structure is presented in (g), and a zoomed-in image of part of the structure is in (f).

Color addition and formation of arbitrary colors is crucial for any display device. Changing the reflectance values of individual pixels contribute to the weighted output. We fabricated a pixelated surface where three different colors are stacked together. The three colors selected in Fig. 4b and 4c are used. Note that although we selected structural colors close to the RGB counterparts of the traditional LED displays, the choice of each structural color and the number of individual blocks are arbitrary. More combinations of colors for pixelated surfaces with varying lateral dimensions can be found in SI (see Fig. Sxx). Our structured surface consisted of 11 X 11 three-pixel units, each consisting of three structural colors next to each other. The individual structural color block dimensions are ≈ 1.5 X 4.5 µm$^2$ forming ≈ 4.5 X 4.5 µm$^2$ three-pixel units. An SEM image of the pixelated surface and a zoomed-in image showing only four units are given in Fig. 4e and Fig. 4h, respectively. The contrast from the SEM image indicates the height difference between individual blocks of the three-pixel unit, which translates to three distinct colors. The optical image of the structured surfaces is given in Fig. 4g, with a zoomed-in image given in Fig. 4f. The optical image confirms that the three-pixel unit is indeed formed by three individual blocks of structural colors with different reflectance states. The next question is, of course, how can this design work dynamically? Utilizing the phase switching of PCMs offer a unique solution for such a problem. One could incorporate the relative reflectance intensity changes from localized structural transitions of PCMs for producing arbitrary colors.[19,23,25,26]

Our work demonstrates the concept of contrast formation and ultrahigh-resolution structured surfaces by producing individually isolated pixel-like structures on a reflective surface. For producing the structures, FIB was used to mill/sputter away materials with a very accurate and controlled procedure involving depth calibration with digital image grayscale values[27]. A reflective heterostructure device, consisting of a 25 nm lossy PCM coating $Sb_2Se_3$ on a gold substrate, was produced and used for the local color contrast formation. By integrating FIB's milling/sputtering power with a precise depth control of a reflective heterostructure, isolated pixel-like structures could be produced. While previously reported works showed reflectance tuning by varying the pitch and separation of individual plasmonic nanopillars, our work demonstrated the addition of height differences between individual pillars creating ultra-high resolution and contrast.

Furthermore, we proposed a metapixel reflective surface where 'primary' structural colors stacked together could produce secondary colors. By varying the reflectance intensity of individual subpixels, color tuning can be achieved mimicking RGB-based display devices like LED displays. PCMs offer a unique advantage for such applications where the relative reflectance is associated with partial crystallization/amorphization of the active material used. The controlled milling/sputtering of materials with FIB and producing local structural colors reduces the task of producing plasmonic nanopillars and resonators with complicated lithography steps. The versatility of FIB for producing such surfaces could provide an advantage where multiple metapixel designs, with different sub-pixel heights and lateral dimensions, can easily be printed on any reflective surface.

## Conclusions

Reflective heterostructure devices, consisting of a low-loss phase change alloy $Sb_2Se_3$ on a reflective gold substrate, were produced using pulsed laser deposited. Pixelated metasurface structures, comprising isolated pillars with relative height variations, were then created using a focused ion beam. We show that contrast can be produced by controlling the depth of each pillar by correlating the milling/sputtering power of the ion beam with a grayscale of a digital image. Although the produced pillars are optically tunable, to move towards a dynamically reconfigurable display device, we proposed a metapixel surface with structural colors stacked together similar to traditional RGB pixelated LED displays. The local reflectance of the reflective heterostructure consisting of phase change alloys can be tuned by a partial crystallization/amorphization of the active material. This can alter the reflectance state of individual pixel elements producing a weighted color. The perfect combination of the switching properties of phase change alloys with pixelated metasurfaces created by FIB could facilitate the move towards dynamically reconfigurable surfaces for display applications.

## Acknowledgments

This project has received funding from the European Union's Horizon 2020 Research and Innovation Programme "BeforeHand" (Boosting Performance of Phase Change Devices by Hetero- and Nanostructure Material Design)" under Grant Agreement No. 824957.

# References

(1) Xiong, K.; Tordera, D.; Jonsson, M. P.; Dahlin, A. B. Active Control of Plasmonic Colors: Emerging Display Technologies. *Reports on Progress in Physics* **2019**, *82* (2), 024501. https://doi.org/10.1088/1361-6633/AAF844.

(2) Xuan, Z.; Li, J.; Liu, Q.; Yi, F.; Wang, S.; Lu, W. Artificial Structural Colors and Applications. *The Innovation* **2021**, *2* (1). https://doi.org/10.1016/j.xinn.2021.100081.

(3) Ko, J. H.; Yoo, Y. J.; Lee, Y.; Jeong, H.-H. H.; Song, Y. M. A Review of Tunable Photonics: Optically Active Materials and Applications from Visible to Terahertz. *iScience* **2022**, *25* (8), 104727. https://doi.org/10.1016/j.isci.2022.104727.

(4) Kim, D. Y.; Kim, M. J.; Sung, G.; Sun, J. Y. Stretchable and Reflective Displays: Materials, Technologies and Strategies. *Nano Convergence 2019 6:1* **2019**, *6* (1), 1–24. https://doi.org/10.1186/S40580-019-0190-5.

(5) Fei Bai, P.; Hayes, R. A.; Jin, M. L.; Shui, L. L.; Yi, Z. C.; Wang, L.; Zhang, X.; Fu Zhou, G. Review of Paper-like Display Technologies. *Progress in Electromagnetics Research* **2014**, *147*, 95–116. https://doi.org/10.2528/PIER13120405.

(6) Hertel, D. Optical Measurement Standards for Reflective E-Paper to Predict Colors Displayed in Ambient Illumination Environments. *Color Res Appl* **2018**, *43* (6), 907–921. https://doi.org/10.1002/COL.22279.

(7) Joo, W. J.; Kyoung, J.; Esfandyarpour, M.; Lee, S. H.; Koo, H.; Song, S.; Kwon, Y. N.; Ho Song, S.; Bae, J. C.; Jo, A.; Kwon, M. J.; Han, S. H.; Kim, S. H.; Hwang, S.; Brongersma, M. L. Metasurface-Driven OLED Displays beyond 10,000 Pixels per Inch. *Science (1979)* **2020**, *370* (6515), 459–463. https://doi.org/10.1126/SCIENCE.ABC8530.

(8) Heydari, E.; Sperling, J. R.; Neale, S. L.; Clark, A. W.; Heydari, E.; Sperling, J. R.; Neale, S. L.; Clark, A. W. Plasmonic Color Filters as Dual-State Nanopixels for High-Density Microimage Encoding. *Adv Funct Mater* **2017**, *27* (35), 1701866. https://doi.org/10.1002/ADFM.201701866.

(9) James, T. D.; Mulvaney, P.; Roberts, A. The Plasmonic Pixel: Large Area, Wide Gamut Color Reproduction Using Aluminum Nanostructures. *Nano Lett* **2016**, *16* (6), 3817–3823.

https://doi.org/10.1021/ACS.NANOLETT.6B01250/ASSET/IMAGES/LARGE/NL-2016-01250B_0006.JPEG.

(10) Badloe, T.; Lee, J.; Seong, J.; Rho, J. Tunable Metasurfaces: The Path to Fully Active Nanophotonics. *Adv Photonics Res* **2021**, *2* (9), 2000205. https://doi.org/10.1002/adpr.202000205.

(11) Franklin, D.; Chen, Y.; Vazquez-Guardado, A.; Modak, S.; Boroumand, J.; Xu, D.; Wu, S. T.; Chanda, D. Polarization-Independent Actively Tunable Colour Generation on Imprinted Plasmonic Surfaces. *Nature Communications 2015 6:1* **2015**, *6* (1), 1–8. https://doi.org/10.1038/ncomms8337.

(12) Zhu, X.; Yan, W.; Levy, U.; Mortensen, N. A.; Kristensen, A. Resonant Laser Printing of Structural Colors on High-Index Dielectric Metasurfaces. *Sci Adv* **2017**, *3* (5). https://doi.org/10.1126/SCIADV.1602487/SUPPL_FILE/1602487_SM.PDF.

(13) Shen, Y.; Rinnerbauer, V.; Wang, I.; Stelmakh, V.; Joannopoulos, J. D.; Soljačić, M. Structural Colors from Fano Resonances. *ACS Photonics* **2015**, *2* (1), 27–32. https://doi.org/10.1021/PH500400W/SUPPL_FILE/PH500400W_SI_001.PDF.

(14) Yang, Z.; Chen, Y.; Zhou, Y.; Wang, Y.; Dai, P.; Zhu, X.; Duan, H. Microscopic Interference Full-Color Printing Using Grayscale-Patterned Fabry–Perot Resonance Cavities. *Adv Opt Mater* **2017**, *5* (10), 1700029. https://doi.org/10.1002/ADOM.201700029.

(15) Hemmatyar, O.; Abdollahramezani, S.; Lepeshov, S.; Krasnok, A.; Brown, T.; Alu, A.; Adibi, A.; Aì, A.; Adibi, A. Advanced Phase-Change Materials for Enhanced Meta-Displays. *null* **2021**. https://doi.org/null.

(16) Koirala, I.; Shrestha, V. R.; Park, C. S.; Lee, S. S.; Choi, D. Y. Polarization-Controlled Broad Color Palette Based on an Ultrathin One-Dimensional Resonant Grating Structure. *Scientific Reports 2017 7:1* **2017**, *7* (1), 1–8. https://doi.org/10.1038/srep40073.

(17) Duan, X.; Kamin, S.; Liu, N. Dynamic Plasmonic Colour Display. *Nature Communications 2017 8:1* **2017**, *8* (1), 1–9. https://doi.org/10.1038/ncomms14606.

(18) Liu, H.; Dong, W.; Wang, H.; Lu, L.; Ruan, Q.; Tan, Y. S.; Simpson, R. E.; Yang, J. K. W. Rewritable Color Nanoprints in Antimony Trisulfide Films. *Sci Adv* **2020**, *6* (51), 7171–7187. https://doi.org/10.1126/sciadv.abb7171.

(19) Cheng, Z.; Milne, T.; Salter, P.; Kim, J. S.; Humphrey, S.; Booth, M.; Bhaskaran, H. Antimony Thin Films Demonstrate Programmable Optical Nonlinearity. *Sci Adv* **2021**, *7* (1), eabd7097. https://doi.org/10.1126/sciadv.abd7097.

(20) Sreekanth, K. V.; Medwal, R.; Srivastava, Y. K.; Manjappa, M.; Rawat, R. S.; Singh, R.; Valiyaveedu Sreekanth, K.; Medwal, R.; Kumar Srivastava, Y.; Manjappa, M.; Singh Rawat, R.; Singh, R.; Sreekanth, K. V.; Medwal, R.; Srivastava, Y. K.; Manjappa, M.; Rawat, R. S.; Singh, R. Dynamic Color Generation with Electrically Tunable Thin Film Optical Coatings. *Nano Lett* **2021**, *21* (23), 10070–10075. https://doi.org/10.1021/acs.nanolett.1c03817.

(21) Abdollahramezani, S.; Hemmatyar, O.; Taghinejad, H.; Krasnok, A.; Kiarashinejad, Y.; Zandehshahvar, M.; Alu, A.; Adibi, A.; Alù, A.; Alù, A.; Adibi, A. Tunable Nanophotonics Enabled by Chalcogenide Phase-Change Materials. *Nanophotonics* **2020**. https://doi.org/10.1515/nanoph-2020-0039.

(22) Wuttig, M.; Bhaskaran, H.; Taubner, T. Phase-Change Materials for Non-Volatile Photonic Applications. *Nat Photonics* **2017**, *11* (8), 465–476. https://doi.org/10.1038/nphoton.2017.126.

(23) Hosseini, P.; Wright, C. D.; Bhaskaran, H. An Optoelectronic Framework Enabled by Low-Dimensional Phase-Change Films. *Nature* **2014**, *511* (7508), 206–211. https://doi.org/10.1038/nature13487.

(24) Vermeulen, P. A.; Yimam, D. T.; Loi, M. A.; Kooi, B. J. Multilevel Reflectance Switching of Ultrathin Phase-Change Films. *J Appl Phys* **2019**, *125* (19), 193105. https://doi.org/10.1063/1.5085715.

(25) Sun, X.; Lotnyk, A.; Ehrhardt, M.; Gerlach, J. W.; Rauschenbach, B. Realization of Multilevel States in Phase-Change Thin Films by Fast Laser Pulse Irradiation. *Adv Opt Mater* **2017**, *5* (12), 1700169. https://doi.org/10.1002/adom.201700169.

(26) Liu, H.; Dong, W.; Wang, H.; Lu, L.; Ruan, Q.; Tan, Y. S.; Simpson, R. E.; Yang, J. K. W. Rewritable Color Nanoprints in Antimony Trisulfide Films. *Sci Adv* **2020**, *6* (51). https://doi.org/10.1126/sciadv.abb7171.

(27) Yimam, D. T.; Liang, M.; Ye, J.; Kooi, B. J. 3D Nanostructuring of Phase-Change Materials Using Focused Ion Beam Towards Versatile Optoelectronics Applications. *Advanced Materials* **2023**, 2303502. https://doi.org/10.1002/ADMA.202303502.

(28) Lee, J. S.; Park, J. Y.; Kim, Y. H.; Jeon, S.; Ouellette, O.; Sargent, E. H.; Kim, D. H.; Hyun, J. K. Ultrahigh Resolution and Color Gamut with Scattering-Reducing Transmissive Pixels. *Nature Communications 2019 10:1* **2019**, *10* (1), 1–9. https://doi.org/10.1038/s41467-019-12689-2.

(29) Kats, M. A.; Blanchard, R.; Genevet, P.; Capasso, F. Nanometre Optical Coatings Based on Strong Interference Effects in Highly Absorbing Media. *Nat Mater* **2013**, *12* (1), 20–24. https://doi.org/10.1038/nmat3443.

Supplementary Information for

# Towards a dynamically reconfigurable pixelated reflective display: Focused ion beam for phase-change metapixel structures

Daniel T. Yimam*, Minpeng Liang, Jianting Ye, Bart J. Kooi*

Zernike Institute for Advanced Materials, University of Groningen, Nijenborgh 4, 9747 AG Groningen, The Netherlands

*Corresponding authors. Email: d.t.yimam@rug.nl, b.j.kooi@rug.nl

# SI 1 – Contrast formation from isolated pixels

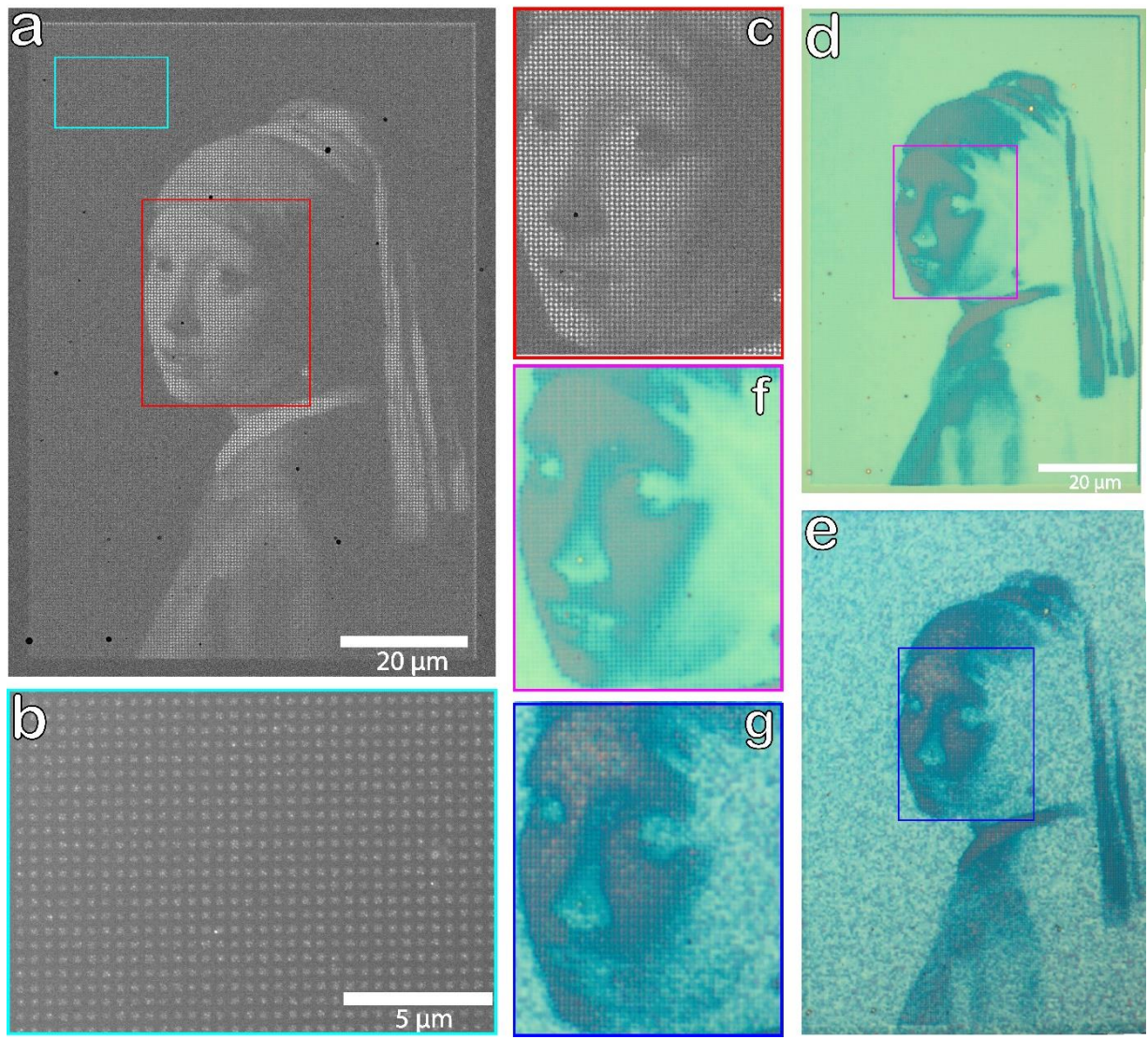

**Figure S1.** Isolated pixelated structures create contrast. An additional example is provided, where the painting *Girl with a Pearl Earring* by Vincent van Gogh is nanoprinted on a reflective surface. (a) An SEM image of the structured surface and closer looks at regions of the pixelated surface in (b) and (c). The optical images of the produced structure for (d, f) the as-deposited amorphous and (e, f) crystalline phases of the $Sb_2Se_3$ layer.

Figure 2 of the main text demonstrates the contrast formation from isolated 'pillar-like' structures. Here, an additional example of a nanostructured surface is presented. The painting *Girl with a Pearl Earring* by Vincent van Gogh is nanoprinted on the surface of our reflective device. Fig. S1(a) presents an SEM image of the structured surface after FIB milling/sputtering. The contrast seen is produced by the backscattered electrons (BSEs) signal, indicating local depth variation. For a better look at formed pillars, two zoomed-in images of some of the regions of the nanostructured surfaces are presented in Fig. S1(b) and (c). Fig. S1(b) shows the individual pillar-like structures. The pillars have the same lateral dimensions and similar depth since the grayscale values in the original uploaded digital images are similar for that specific region. However, for the region presented in Fig. S1(c), the largest grayscale value difference in the digital image and relatively large depth variations between isolated pillars can be observed. The relative depth variations are responsible for the contrast formation.

The relative depth contrast between pillar-like structures in the SEM image also translates into an optical contrast. The optical images of the structured metasurface for the as-deposited amorphous and crystalline phase of the active $Sb_2Se_3$ layer are presented in Fig. S1(d) and (e), respectively. The original image is produced on the surface of the reflective device, and the color contrast seen in Fig. S1(d) is due to how deep each pillar-like structure is throughout the thickness of the film. Moreover, the phase transition of the active $Sb_2Se_3$ layer creates another set of colors and thus contras, as seen in Fig. S1(e). Optical images of zoomed-in regions in the nanostructured surface for the as-deposited amorphous and crystalline phases of the $Sb_2Se_3$ layer are presented in Fig. S1(f) and (g), respectively. Each pillar-like structure is resolvable in both images, and local color variation can be seen clearly.

## SI 1 – Pixelated metasurface from color combinations

Creating local structural colors is one of the many advantages of using FIB to nanostructure reflective surfaces. We showed that this capability could produce a metapixel surface where different structural colors could be stacked together, similar to traditional emissive-based LED pixels where RGB combinations are used. The main text presented a metapixel design where a reflective RGB combination was demonstrated as a proof of concept. Here, additional color combinations are given. The idea here is that although we can perfectly imitate traditional RGB pixelated display devices, we can extend the capabilities with other structural colors. Since the intensity variation is due to the phase-switching of the active $Sb_2Se_3$ layer, the contrast could be, in principle, produced from any color combinations. The ability to produce metapixels from any structural color combinations might find application outside the traditional RGB color combinations.

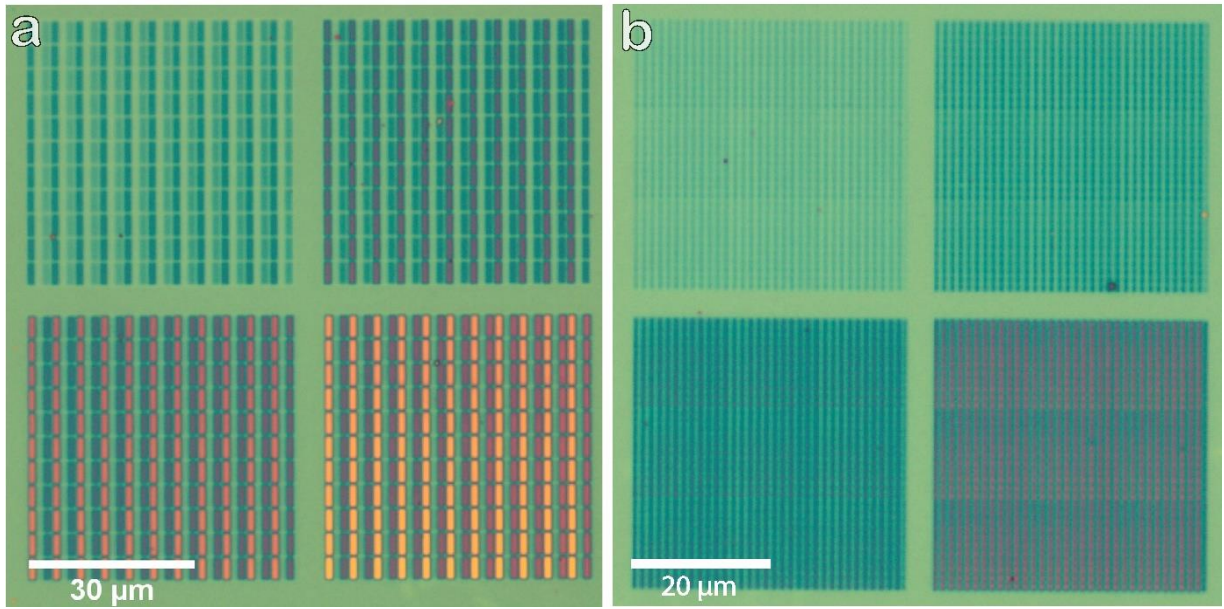

**Figure S2**. Structural color combinations for pixel formation. Additional pixelated surfaces are given in (a), where various metapixel designs from a combination of different structural colors are seen. In (b), the flexibility of using FIB to create metapixels is demonstrated by demonstrating the production of different metapixels of reduced pixel sizes and various color combinations.

Another crucial advantage of using FIB to nanostructure metasurfaces is the great control over lateral and depth dimensions when milling/sputtering material away. When moving towards a reflective-based pixelated surface, dimensional control is crucial. Vivid contrasts and high-resolution image formation require a reduction in pixel sizes. In Fig. S2(b), we showed our ability to produce high pixel density metasurface by combining smaller dimension structural colors. Similar to what has been demonstrated in Fig. S2(a), different color combinations can be produced. The number of pixels packed in each of the structures seen in Fig. S2(b) is nine times higher than the individual structures seen in (a).